\documentclass[onecolumn,showpacs,amsmath,amssymb,superscriptaddress,]{revtex4}

\usepackage{amssymb}

\usepackage{graphicx}
\usepackage{epsfig}
\usepackage{subfigure}
\usepackage{amsmath,amssymb}

\newcommand{\be}{\begin{equation}}
\newcommand{\ee}{\end{equation}}
\newcommand{\ba}{\begin{eqnarray}}
\newcommand{\ea}{\end{eqnarray}}
\newcommand{\ban}{\begin{eqnarray*}}
\newcommand{\ean}{\end{eqnarray*}}
\newcommand{\ket}[1]{\mbox{$ | #1 \rangle $}}
\newcommand{\bra}[1]{\mbox{$ \langle #1 | $}}

\newcommand{\one}{\leavevmode\hbox{\small1\normalsize\kern-.33em1}}

\newcommand{\headtitle}{\fontsize{18pt}{\baselineskip}\selectfont}

\begin{document}

\title{\headtitle\textbf{Correlations and thermalization in driven cavity arrays}}
%%%%alphabetical listing

\author{Li Dai}\email{daili@nus.edu.sg}
\affiliation{Centre for Quantum Technologies, National University of
Singapore, 3 Science Drive 2, Singapore 117543}
\address{Department of Physics, National University of
Singapore, 2 Science Drive 3 Singapore 117542}

\author{Dimitris G. Angelakis}
\affiliation{Centre for Quantum Technologies, National University of
Singapore, 3 Science Drive 2, Singapore 117543} \affiliation{Science
Department, Technical University of Crete, Chania, Crete, Greece,
73100, EU}

\author{Leong Chuan Kwek}
\affiliation{Centre for Quantum Technologies, National University of
Singapore, 3 Science Drive 2, Singapore 117543}
\affiliation{National Institute of
      Education and Institute of Advanced Studies,
      Nanyang Technological University, 1 Nanyang Walk, Singapore
      637616}

\author{S. Mancini}
\affiliation{School of Science and Technology, University of
Camerino, 62032 Camerino, Italy EU}

\begin{abstract}
\noindent\textbf{Abstract.} We show that long-distance steady-state
quantum correlations (entanglement) between pairs of cavity-atom
systems in an array of lossy and driven coupled resonators can be
established and controlled. The maximal of entanglement for any pair
is achieved when their corresponding direct coupling is much smaller
than their individual couplings to the third party. This effect is
reminiscent of the coherent trapping of the $\Lambda-$type
three-level atoms using two classical coherent fields. Different
geometries for coherent control are considered. For finite
temperature, the steady state of the coupled lossy atom-cavity
arrays with driving fields is in general not a thermal state. Using
an appropriate distance measure for quantum states, we find that the
change rate of the degree of thermalization with respect to the
driving strength is consistent with the entanglement of the system.

\end{abstract} 

\maketitle

\section{\large introduction}
 Coupled cavity arrays have recently been proposed
 as a novel system for realizing quantum computation
\cite{angelakis-ekert04} and for simulations of quantum many-body
systems \cite{simulation of many body system}. More recently, the steady-state
polaritonic \cite{two-state} and membrane entanglement
\cite{ple-hue-har}     of
driven cavity arrays were studied under realistic dissipation environment. Also, there has been 
an attempt to relate coupled cavity arrays with Josephson oscillations \cite{coherent control of photon emission}.

At finite temperature, it is expected that the steady state of the
coupled-cavity system is a thermal state, since standard statistical
mechanics tells us that if a system interacts with a large reservoir
at a fixed temperature, it will relax eventually to an equilibrium
state characterized by the Boltzmann distribution with a
well-defined temperature, i.e. that of the reservoir. However, such thermal 
relaxation  is true only for some simple systems such as a single
empty cavity coupled to a thermal bath \cite{Carmichael}. For many
other systems e.g. coupled cavities with external pumping lasers,
the steady state does not need to be a thermal state, and its
deviation from a thermal state depends on various factors:
inter-cavity couplings, presence of the pump, detuning and so forth.

The purpose of this article is twofold: firstly we wish to demonstrate the
possibility of achieving coherent control of the steady-state
entanglement between mixed light-matter excitations generated in
macroscopically separated atom-cavity systems, and secondly, we hope to
elucidate the conditions under which the steady state differs from a
thermal state, especially the relation between the thermalization of
the system and the correlations of the subsystems, using the coupled
atom-cavity system as an example. This paper is organized as
follows. In Sec. II, we introduce the setup and the Hamiltonian for
coherent control of the steady-state entanglement. In Sec. III, we
derive an effective equation for the dynamics of the system. In Sec.
IV, we discuss the coherent control of the steady-state
entanglement. In Sec. V, we discuss an alternative setup: two
coupled cavities with three driving fields. In Sec. VI, we discuss
the thermalization of two defect cavities coupled to one driven wave
guide in between. In Sec. VII, we summarize our result.

\section{\large The setup and the
Hamiltonian}\label{chapter3-3cavity-setup-Hamiltonian}

\begin{figure}[t]
\epsfxsize=.30\textwidth \epsfysize=.25\textwidth
\centerline{\epsffile{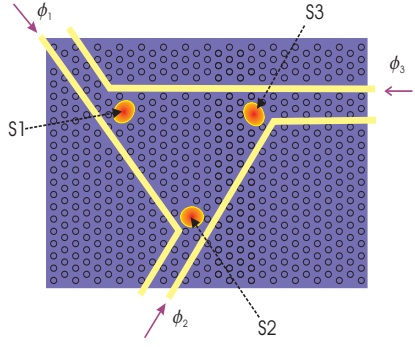}}
%\vspace*{5mm}
\caption[Schematic representation of three interacting cavity-atom
systems]{(color online).  Schematic representation of three
interacting cavity-atom systems ($S_{1}$, $S_{2}$, $S_{3}$) based on
a possible implementation using photonic crystals (for illustration
purposes only): the connecting wave guides carrying the driving
classical fields with phases $\phi_{1}$, $\phi_{2}$, $\phi_{3}$ are
replaced by fibers or stripline microresonators for different
implementations \cite{pbgs,rest}. The three wave guides and three
driving fields are labeled with the same indices to the phases
$\phi_{1}$, $\phi_{2}$, $\phi_{3}$. } \label{Fig.3-cav}
\end{figure}

The setup we study is shown in Fig. \ref{Fig.3-cav}. It contains
three interacting atom-cavity systems ($S_{1}$, $S_{2}$, $S_{3}$)
connected by three waveguides/fibres. Each waveguide/fibre is pumped
by a classical field with a phase $\phi_{i}$, ($i=1,2,3$). The setup
could be realized in a variety of cavity-quantum-electrodynamics
(cavity-QED) technologies including photonic crystals, circuit QED,
toroidal cavities connected through fibers, Fabry-Perot cavities and
coupled defect cavities interacting with quantum dots
\cite{pbgs,rest}. Light from the connecting waveguides/fibers can
directly couple to the photonic modes of the atom-cavity systems
through tunneling or evanescent coupling. In each atom-cavity site
we assume the interaction and the corresponding nonlinearity to be
strong enough with at most one excited polariton\cite{two-state}.

The Hamiltonian describing the system is
\begin{align}\label{H-original-1} H_{0}=&H_{a,0}+H_{p,0}+H_{J,0},\\
H_{a,0}=&\sum_{i=1}^{3}\omega_{c,i}a_{i}^{\dagger}a_{i},\,\,\,H_{p,0}=\sum_{i=1}^{3}\omega_{p,i}P_{i}^{\dagger}P_{i},\end{align}
\begin{align}
H_{J,0}=\sum_{i=1}^{3}J_{i}(a_{i}^{\dagger}(P_{i}+P_{i+1})+a_{i}(P_{i}^{\dagger}+P_{i+1}^{\dagger}))+\sum_{i=1}^{3}(\alpha_{i}e^{i(\phi_{i}-\omega_{d}t)}a_{i}^{\dagger}+\alpha_{i}e^{-i(\phi_{i}-\omega_{d}t)}a_{i}),\end{align}
where $H_{a,0}$ and $H_{p,0}$ are the free Hamiltonians of the wave
guides and cavities, with $a_{i}^{\dagger}$, $a_{i}$ the field
operators of the single-mode wave guides and $\omega_{c,i}$
($\omega_{p,i}$) the frequencies of $i$th waveguide mode (the
polariton in $i$th cavity). $P_{i}^{\dagger}$ ($P_{i}$) the
operators describing the creation (annihilation) of a mixed
atom-photon excitation (polariton) at the $i$th cavity-atom system
($P_{4}\equiv P_{1}$). The first summation in $H_{J,0}$ describes
couplings between cavities and wave guides, with $J_{i}$ the
coupling strength between the photon mode in the $i$th waveguide and
the adjacent two polaritons. The second summation in $H_{J,0}$
describes the classical driving of the wave guides, where
$\alpha_{i}$ is proportional to the amplitude of the $i$th driving
field with $\phi_{i}$ its phase and $\omega_{d}$ the frequency of
the driving fields.

It can be seen that the Hamiltonian $H_{0}$ in Eq.
(\ref{H-original-1}) is explicitly time-dependant. To remove the time dependence, we make the following
transformation \cite{rotating-frame}.\begin{align}
H=U_{1}^{\dagger}H_{0}U_{1}-iU_{1}^{\dagger}\frac{\partial
U_{1}}{\partial t},\end{align} where
$U_{1}=e^{-it\omega_{d}(\sum_{i=1}^{3}(a_{i}^{\dagger}a_{i}+P_{i}^{\dagger}P_{i})}$.
After a straightforward calculation, we obtain
\begin{align}\label{H-original}
H&=H_{a}+H_{p}+H_{J},\\
H_{a}&=\sum_{i=1}^{3}(\omega_{c,i}-\omega_{d})a_{i}^{\dagger}a_{i},\,\,\,H_{p}=\sum_{i=1}^{3}(\omega_{p,i}-\omega_{d})P_{i}^{\dagger}P_{i},\label{chapter2-Hamiltonian-empty-doped-cavity}\\
H_{J}&=\sum_{i=1}^{3}J_{i}(a_{i}^{\dagger}(P_{i}+P_{i+1})+a_{i}(P_{i}^{\dagger}+P_{i+1}^{\dagger}))+\sum_{i=1}^{3}(\alpha_{i}e^{i\phi_{i}}a_{i}^{\dagger}+\alpha_{i}e^{-i\phi_{i}}a_{i}).\label{chapter2-Hamiltonian-interaction-driving}\end{align}
The density matrix $\rho(t)$ of the system associated with $H$ is
related, in the following way, to the density matrix $\rho_{0}(t)$
of the system associated with $H_{0}$. \ba
\rho(t)=U_{1}^{\dagger}\rho_{0}(t)U_{1}.\ea We say that the new
Hamiltonian $H$ is written in the rotating frame of the driving
lasers.

\section{\large The dynamics of the
system}\label{chapter3-3cavity-master-equation} In this section, we
will derive the dynamical equation for the system.

The polaritons and waveguide modes in our system described in the
last section are assumed to decay with rates $\gamma$ and $\kappa$
respectively. The master equation for the whole system density
operator $R$ is: \begin{align}
\frac{dR}{dt}=L_{a}R+L_{p}R+L_{J}R,\label{me}\end{align}
\begin{align}\label{super operator 1}
L_{a}R&=-i[H_{a}\,,R]+L_{a}^{\prime}R,\\\label{super operator
2}L_{p}R&=-i[H_{p}\,,R]+L_{p}^{\prime}R,\\\label{super operator
3}L_{J}R&=-i[H_{J}\,,R],\end{align} where $H_{a}$, $H_{p}$ and
$H_{J}$ are given by Eqs.
(\ref{chapter2-Hamiltonian-empty-doped-cavity}), and
(\ref{chapter2-Hamiltonian-interaction-driving}) respectively, and
\begin{align}
L_{a}^{\prime}R=\frac{\kappa}{2}\sum_{i=1}^{3}(2a_{i}Ra_{i}^{\dagger}-a_{i}^{\dagger}a_{i}R-Ra_{i}^{\dagger}a_{i}),\,\,\,L_{p}^{\prime}R=\frac{\gamma}{2}\sum_{i=1}^{2}(2\sigma_{i}R\sigma_{i}^{\dagger}-\sigma_{i}^{\dagger}\sigma_{i}R-R\sigma_{i}^{\dagger}\sigma_{i}).
\end{align}

We use the projection operator method in Ref. \cite{two-state}.
To this end, we define the projector $PR=r_{ss}\otimes
\textrm{tr}_{a_{1},a_{2},a_{3}}R$, where $r_{ss}$ satisfying
$L_{a}r_{ss}=0$ is the equilibrium state of the three wave guides,
which is close to the vacuum state $\ket{000}\bra{000}$ when weak
driving for the wave guides is assumed i.e. $\alpha_{i}\le
J_{i}\ll\kappa$ ($i=1,2,3$). The orthogonal complement of $P$ is
$Q=1-P$. The operators $P$ and $Q$ have the properties that
\cite{projection-operator-method} \ba\label{property 1}
PL_{p}&=&L_{p}P\,,\\\label{property
2}PL_{a}&=&L_{a}P=0,\\\label{property 3}PL_{J}P&=&0.\ea Applying $P$
and $Q$ respectively to Eq. (\ref{me}) and using the properties
(\ref{property 1}), (\ref{property 2}) and (\ref{property 3}), we
get \ba\label{p}
P\frac{dR}{dt}&=&PL_{p}PR(t)+PL_{J}QR(t),\\\label{q}Q\frac{dR}{dt}&=&Q(L_{a}+L_{p}+L_{J})QR(t)+QL_{J}PR(t).\ea
Formally integrate (\ref{q}) to get\ba
QR(t)=\int_{-\infty}^{t}Qe^{(L_{a}+L_{p}+L_{J})(t-t^{\prime})}L_{J}PR(t^{\prime})dt^{\prime},\ea
which is then replaced into Eq. (\ref{q}). For the case
$J_{i}\ll\kappa$, ($I=1,2,3$) we only keep the second order in
$J_{i}$\,. By tracing out $a_{1}$, $a_{2}$ and $a_{3}$, we obtain
\ba \frac{d\rho}{dt}&=&-i[H_{p}\,,\rho(t)]+L_{p}^{\prime}\rho(t)\nonumber\\
&+&\int_{0}^{\infty}dt^{\prime}\textrm{tr}_{a_{1},a_{2},a_{3}}[{L_{J}e^{(L_{a}+L_{p})t^{\prime}}}L_{J}e^{-L_{p}t^{\prime}}(r_{ss}\otimes
\rho)]\ea Substituting $L_{a}$, $L_{p}$ and $L_{J}$ with expressions
(\ref{super operator 1}), (\ref{super operator 2}) and (\ref{super
operator 3}), we get \ba\label{eq-eff}
\frac{d\rho}{dt}=&-&i[H_{\mbox{\rm
eff}}\,,\rho]+\sum_{i=1}^{3}(\Gamma_{i-1}z_{i-1}+\Gamma_{i}z_{i})F_{i,i}^{P}\rho\nonumber\\
&+&\sum_{i=1}^{3}\Gamma_{i}(F_{i,i+1}^{P}\rho+F_{i+1,i}^{P}\rho)\,,\ea
with $\displaystyle H_{\mbox{\rm
eff}}=\sum_{i=1}^{3}(\omega_{p,i}-\omega_{d})P_{i}^{\dagger}P_{i}
+\sum_{i=1}^{3}\Gamma_{i}y_{i}(P_{i}^{\dagger}P_{i+1}+P_{i}^{\dagger}P_{i+1})
+\sum_{i=1}^{3}(\Gamma_{i}y_{i}P_{i}^{\dagger}P_{i+1}$
+$\Gamma_{i}x_{i}(P_{i}^{\dagger}+P_{i+1}^{\dagger}))+h.c.\,$,$\vspace*{1mm}$
where $h.c.$ denotes the Hermitian conjugation of its previous
summation. The first two summations in $\displaystyle H_{\mbox{\rm
eff}}$ cancel with each other with a proper choice of
$\omega_{p,i}$\,. $F_{i,j}^{P}(\rho)=2P_{i}\rho
P_{j}^{\dagger}-P_{i}^{\dagger}P_{j}\rho-\rho
P_{i}^{\dagger}P_{j}\,$, $\displaystyle
\Gamma_{i}=2J_{i}^{2}\kappa/(\kappa^{2}+4\Delta_{i}^{2})$,$\vspace*{2mm}$
$x_{i}=-\alpha_{i}e^{i\phi_{i}}(2\Delta_{i}+i\kappa)/(J_{i}\kappa)$,
$y_{i}=-2\Delta_{i}/\kappa$,
$\Delta_{i}=\omega_{c,i}-(\omega_{p,i}+\omega_{p,i+1})/2$,
$\omega_{p,4}\equiv \omega_{p,1}$, $z_{i}=1+\gamma/(4\Gamma_{i})$,
$\Gamma_{0}\equiv\Gamma_{3}$ and $z_{0}\equiv z_{3}$. It can be seen
from Eq. (\ref{eq-eff}) that the couplings and detunings between the
wave guide and its adjacent two polaritons induce an effective
interaction between them given by $\Gamma_{i}y_{i}$ (see
$H_{\textrm{eff}}$). The driving on the wave guides is equivalently
transferred to the driving on the polaritons ($\Gamma_{i}x_{i}$ in
$H_{\textrm{eff}}$), which decay with rates
$\Gamma_{i-1}z_{i-1}+\Gamma_{i}z_{i}=\Gamma_{i-1}+\Gamma_{i}+\gamma$.
Since $\Gamma_{i}$ is related to $\kappa$, the polaritons
effectively have two different channels for the decay. They can either decay directly to the
surrounding with $\gamma$ and they can also dissipate energy via the coupling $J_{i-1}$ or
$J_{i}$ ($J_{0}\triangleq J_{3}$) to the adjacent two leaky wave
guides (who also decay by $\kappa$). We notice that the second
channel also mixes the polaritons' operators, as  seen in the
second line of Eq. (\ref{eq-eff}). This mixing is actually one of the main reasons 
for entanglement creation among the polaritons. Note that the
other two contributing factors are the interactions among polaritons
and the driving on them.

\section{\large Coherent control of the steady-state entanglement}\label{chapter3-3cavity-steady-state}

We now derive the steady state $\rho_{ss}$ by requiring that
$\displaystyle\frac{d\rho_{ss}}{dt}=0$ in Eq. (\ref{eq-eff}). This
is done numerically due to the large number of coupled equations
involved. For a three-polariton density matrix, we trace
out the polaritonic degree of freedom of cavity 1 and calculate the
polaritonic entanglement of formation between cavity 2 and 3 using
the concurrence as a measure \cite{Woot}. The concurrence
$C(\rho_{ss})$ is effectively a function of the parameters
$x_{i}$\,, $y_{i}$ and $z_{i}$. We perform a numerical optimization
of $C(\rho_{ss})$ by varying these parameters and find that
$C(\rho_{ss})$ is larger when $\Gamma_{2}\ll\Gamma_{1}=\Gamma_{3}$
and $x_{3}=-x_{1}$, i.e. the first and third driving fields have
equal intensity but opposite phases. We also note here that the
relation $\Gamma_{2}\ll \Gamma_{1}=\Gamma_{3}$ indicates that the
coupling between the two cavities in question is much weaker than
the coupling between each one of the cavities and the third cavity. Also the state
of the polariton in cavity 1 for the maximum entanglement point is
found to be almost a pure state at ground energy level and therefore
almost uncorrelated to the polaritons in cavity 2 and 3. Thus, the
total density matrix $\rho\approx
\ket{\textrm{ground}}\bra{\textrm{ground}}\otimes \rho_{2,3}$.
Although this result initially looks counter-intuitive, it can be
explained as follows: the maximum entanglement between the two
parties, i.e. cavities 2 and 3, in a three-party system, is attained
when the state of the third party, i.e. cavity 1, nearly factorizes
in the combined three-party state. The fact that this is happening
for relatively strong couplings of $J_{12}\equiv J_1$ and $J_{13}
\equiv J_3$ compared to $J_{23}\equiv J_2$ is reminiscent of the
behavior of a coherent process taking place. It is interesting to 
observe an analogy here with the case of coherently superposing two
initially uncoupled ground states in a $\Lambda$-type quantum system
through an excited state using two classical fields to mediate the
interaction \cite{Scully, EIT-Harris}.

\begin{figure}[h]
\epsfxsize=.25\textwidth \epsfysize=.2\textwidth
\centerline{\epsffile{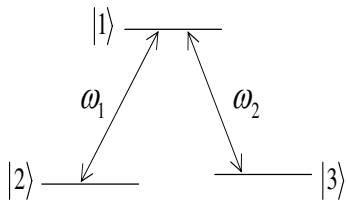}}
%\vspace*{5mm}
\caption[Coherent trapping of correlations in a 3-cavity
system]{(color online).  The coherent trapping of a $\Lambda$-type
three-level atom driven by two classical fields on resonance, where
$\omega_{1}$ and $\omega_{2}$ are the frequencies of the two driving
fields . If the states $\ket{2}$ and $\ket{3}$ are degenerate, one
could use two laser fields with different polarizations to
distinguish the two driving paths ($\ket{2}$ to $\ket{1}$, and
$\ket{3}$ to $\ket{1}$).} \label{chapter3-coherent-trapping}
\end{figure}

In figure \ref{chapter3-coherent-trapping}, we compare our setup for
entanglement control of three-coupled-cavity system with the
coherent population trapping in a three-level atom. For the latter,
if the two driving fields have opposite phases and the atom's
initial state is $(\ket{2}+\ket{3})/\sqrt{2}$, there will be no
population in the excited state $\ket{1}$ and the atom will remain
in a superposition of the states $\ket{2}$ and $\ket{3}$. Note 
that the states $\ket{2}$ and $\ket{3}$ are not coupled in this case. The
superposition of them is established by a quantum interference in
the state $\ket{1}$ \cite{Scully}. It appears that the quantum
correlation in our setup is somewhat  "trapped" in the cavity 2 and cavity 3
if the driving fields 1 and 3 have opposite phases (The cavity 2 and
cavity 3 are almost uncoupled. In this case, it is numerically
verified that the driving field between them has almost no influence
on their steady-state entanglement).

\begin{figure}
\epsfxsize=.35\textwidth \epsfysize=.3\textwidth
\centerline{\epsffile{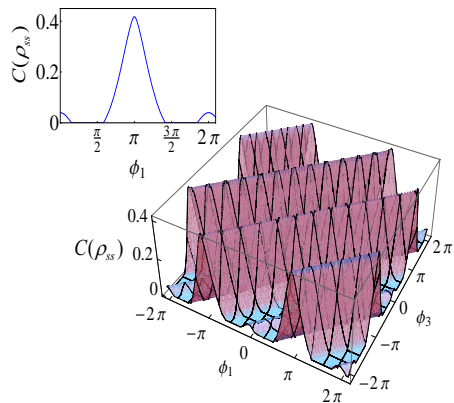}}
%\vspace*{5mm}
\caption[The coherent effect of the entanglement in a 3-cavity
system]{(color online).  The concurrence between the polaritons in
cavity 2 and cavity 3 as a function of $\phi_{1}$ and $\phi_{3}$.
$x_{1}=1.67e^{i\phi_{1}}$, $x_{3}=1.67e^{i\phi_{3}}$. When
$\phi_{1}-\phi_{3}=(2k+1)\pi$ ($k$ is an integer), the concurrence
reaches a maximum of 0.417. The upper left figure is the sectional
view at $\phi_{3}=0$.} \label{Fig.phase13}
\end{figure}

The observation in the above paragraph is further justified by
noticing that $C(\rho_{ss})$ is varied with the phases of the first
and third driving fields. In Fig. \ref{Fig.phase13} we plot
$C(\rho_{ss})$ as a function of the phases of driving fields with
$z_{1}=z_{3}=1.01$ and $z_{2}=11$. When the phase difference is
$\phi_{1}-\phi_{3}=(2k+1)\pi$ ($k$ is an integer), we get a maximum
of 0.417. For general phase relations, an oscillatory behavior
characteristic of the expected coherent effect takes place. There is
a corresponding oscillatory behavior for the $\Lambda$-type
three-level atom: the summation of the modulus square of the
amplitudes in the states $\ket{2}$ and $\ket{3}$ is a periodic
function of the phase difference between the two driving fields and
takes a maximum when their phases are opposite \cite{Scully}.

\section{\large An alternative setup: Two coupled cavities with three
driving fields}

\begin{figure}
\epsfxsize=.3\textwidth \epsfysize=0.25\textwidth
\centerline{\epsffile{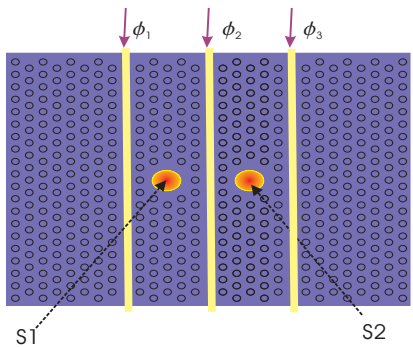}} \caption[Schematic diagram of
the two cavity-atom systems]{(color online).  Schematic diagram of
the two coupled defect cavities in which there are three wave guides
carrying the three respective classical laser fields. Note that each
waveguide carrying classical fields can also be replaced by fibers
or stripline microresonators for different implementation
technologies \cite{pbgs,rest}.} \label{Fig.2-cav}
\end{figure}

In Section \ref{chapter3-3cavity-steady-state}, we find that when
the entanglement between the two of the three cavities reaches a
maximum value, the third cavity nearly decouples from the two
cavities. It therefore seems that the third cavity plays absolutely no role in the
establishment of the entanglement between the other two cavities. To
check if this argument is correct and identify the  role of the
third cavity in the entanglement generation and control, we
remove the third cavity and investigate the entanglement of the
remaining two cavities. This new setup is shown in Fig. \ref{Fig.2-cav},
where there are three wave guides coupled to two cavity-atom systems
and these three wave guides are driven by three classical fields
respectively. We analyze the polaritonic entanglement between
cavity 2 and 3 (relabeled as $S_{1}$ and $S_{2}$ in Fig.
\ref{Fig.2-cav}).

The Hamiltonian and the derivation of the effective master equation
are similar to those for the three-cavity setup in Section
\ref{chapter3-3cavity-setup-Hamiltonian} and
\ref{chapter3-3cavity-master-equation}. We therefore omit the detailed derivation
steps and provide only the final effective master equation.
 \ba\label{eq-eff-2}
\frac{d\rho}{dt}=&-&i[H_{\mbox{\rm eff}}'\,,\rho]\nonumber\\
&+&(\Gamma_{2}z_{2}+\Gamma_{1})F_{1,1}^{P}\rho+(\Gamma_{2}z_{2}+\Gamma_{3})F_{2,2}^{P}\rho\nonumber\\
&+&\Gamma_{2}(F_{1,2}^{P}\rho+F_{2,1}^{P}\rho)\,,\ea with
$\displaystyle H_{\mbox{\rm
eff}}'=(\Gamma_{2}y_{2}P_{1}^{\dagger}P_{2}$
+$\sum_{i=1}^{2}(\Gamma_{i}x_{i}+\Gamma_{i+1}x_{i+1})P_{i}^{\dagger})+h.c.\,$,$\vspace*{1mm}$
where $h.c.$ denotes the Hermitian conjugation of its previous
summation. $F_{i,j}^{P}(\rho)$ is defined in Section
\ref{chapter3-3cavity-master-equation} as $2P_{i}\rho
P_{j}^{\dagger}-P_{i}^{\dagger}P_{j}\rho-\rho
P_{i}^{\dagger}P_{j}\,$, $\displaystyle
\Gamma_{i}=2J_{i}^{2}\kappa/(\kappa^{2}+4\Delta_{i}^{2})$,$\vspace*{2mm}$
$x_{i}=-\alpha_{i}e^{i\phi_{i}}(2\Delta_{i}+i\kappa)/(J_{i}\kappa)$,
$y_{2}=-2\Delta_{2}/\kappa$, $\Delta_{1}=\omega_{c,1}-\omega_{p,1}$,
$\Delta_{2}=\omega_{c,1}-(\omega_{p,1}+\omega_{p,3})/2$,
$\Delta_{3}=\omega_{c,1}-\omega_{p,3}$,
$z_{i}=1+\gamma/(4\Gamma_{i})$.

\begin{figure}[t]
\epsfxsize=.35\textwidth \epsfysize=0.3\textwidth
\centerline{\epsffile{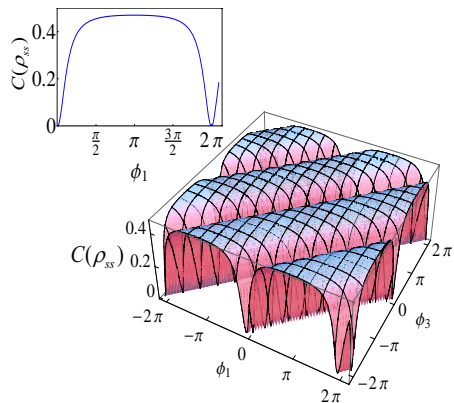}}
%\vspace*{5mm}
\caption[The coherent effect of the entanglement in a 2-cavity
system]{(color online).  The concurrence between two cavities
-Fig.\ref{Fig.2-cav}- as a function of $\phi_{1}$ and $\phi_{3}$.
$x_{2}=y_{2}=0$, $x_{1}=5e^{i\phi_{1}}$,$x_{3}=5e^{i\phi_{3}}$,
$\Gamma_{1}=\Gamma_{3}=1.316\times 10^{8}$ and $\Gamma_{2}=10^{10}$.
When $\phi_{1}-\phi_{3}=(2k+1)\pi$ ($k$ is an integer), the
concurrence reaches a maximum of 0.470. The upper left figure is the
sectional view at $\phi_{3}=0$.} \label{Fig.phase23}
\end{figure}

The optimization of this entanglement gives similar values of the
parameters like the ones used
 above except that the values for $\Gamma_{i}$ are reversed, i.e.
 $\Gamma_{2}\gg\Gamma_{1}=\Gamma_{3}$; however,
 the concurrence reaches a maximum of 0.47. Again the
dependence $\phi_{1}-\phi_{3}=(2k+1)\pi$ ($k$ is an integer) is
apparent (see Fig. \ref{Fig.phase23}). However, if we compare the
 insets in Fig. \ref{Fig.phase13} and Fig. \ref{Fig.phase23} for the
 cross-sectional plots of the concurrence for $\phi_{3}=0$, we
 see that the plot in Fig. \ref{Fig.phase13} has a narrower peak
 whereas the plot in Fig. \ref{Fig.phase23} is broader. This implies
 that the maximum concurrence for configuration in Fig.
 \ref{Fig.2-cav} is substantially more stable against variation in
 the phases $\phi_{1}$ and $\phi_{3}$ than that in Fig.
 \ref{Fig.3-cav}. However, when the dissipation (parametrized by $\gamma$ in $z_{i}$)) increases,
 the entanglement in the latter configuration decreases more slowly than
the former one. This can be numerically verified. Thus we conclude
that cavity 1 in Fig. \ref{Fig.3-cav} not only mediates coherently 
between cavities 2 and 3, but it also stabilizes the amount of
entanglement between the two cavities.

\begin{figure}[h]
\begin{center}
%\vspace{10cm}
\includegraphics[width=0.25\textwidth,height=0.2\textwidth]{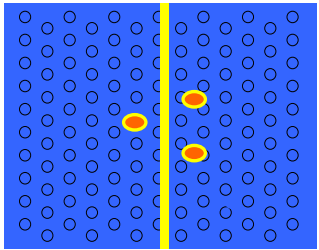}
\end{center}
\caption{(color online).  Three defect cavities coupled to one wave
guide.}\label{more-cavity-2}
\end{figure}

There are many other configurations for the coupled-cavity setup.
For instance, one could consider an extension of the setup in Ref.
\cite{two-state} to three defect cavities, as shown in Fig.
\ref{more-cavity-2}. However, numerical optimization for this
extension and many others does not seem to increase the polaritonic
entanglement between any two cavities. Therefore, the setups in Fig.
\ref{Fig.3-cav} and \ref{Fig.2-cav} appear to be optimal ones for
two-polariton entanglement.

\section{\large Thermalization of the coupled-cavity system}\label{model}

\begin{figure}[h]
\begin{center}
%\vspace{10cm}
\includegraphics[width=0.25\textwidth,height=0.2\textwidth]{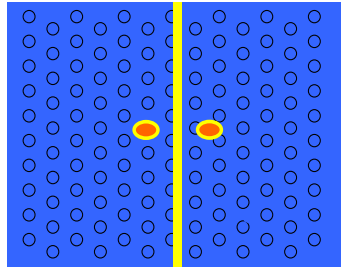}
\end{center}
\caption{(color online).  Two defect cavities coupled to one wave
guide.}\label{2-cavity}
\end{figure}

In this section, we consider the thermalization of the lossy driven atom-cavity
system. For simplicity, we consider a simpler system which involves
two defect cavities coupled to a driven wave guide, as shown in Fig.
\ref{2-cavity}.

This system was studied in Ref. \cite{two-state}, where the
reservoir temperature is set to be zero and an analytical solution
was obtained (see Eq. (23)-(29) therein). For finite temperature,
the master equation needs to be modified, i.e. Eq. (12) and (13) of
Ref. \cite{two-state} are replaced by
\begin{align}
L_{a}'R=&\kappa(n_{c}+1)(2aRa^{\dagger}-a^{\dagger}aR-Ra^{\dagger}a)+\kappa
n_{c}(2a^{\dagger}Ra-aa^{\dagger}R-Raa^{\dagger}),\\
L_{p}'R=&\sum_{i=1}^{2}\gamma(n_{p}+1)(2\sigma_{i} R
\sigma^{\dagger}_{i}-\sigma^{\dagger}_{i}\sigma_{i}
R-\rho\sigma^{\dagger}_{i}\sigma_{i})+\gamma
n_{p}(2\sigma^{\dagger}_{i} R \sigma-\sigma\sigma^{\dagger}_{i}R-R
\sigma_{i}\sigma^{\dagger}_{i}),\label{polariton decay}\end{align}
where $n_{c}=\frac{1}{e^{\hbar\omega_{cav}/k_{B}T_{R}}-1}$ is the
mean photon number at the reservoir temperature $T_{R}$ and the
cavity frequency $\omega_{cav}$. Similarly
$n_{p}=\frac{1}{e^{\hbar\omega_{pol}/k_{B}T_{R}}-1}$ is the mean
photon number at the reservoir temperature $T_{R}$ and the
polaritonic frequency $\omega_{pol}$.

The effective master equation for the two polaritons can be obtained
using the same method in Ref. \cite{two-state}. For
$T_{R}\ll\hbar\omega_{pol}/k_{B}$, the temperature terms in Eq.
(\ref{polariton decay}) are preserved in the final effective master
equation i.e. Eq. (20) of Ref. \cite{two-state}. The steady
state $\rho^{ss}$ is obtained by requiring
$\frac{d\rho^{ss}}{dt}=0$. To characterize the degree of
thermalization of the steady state, we calculate the distance
between the steady state and a thermal state, using the following
distance measure \cite{trace distance}:
\begin{align}\label{distance measure}
d(\rho^{ss},\,\,\rho^{th})=\frac{1}{2}\textrm{tr}|\rho^{ss}-\rho^{th}|.\end{align}
The trace distance $d(\rho^{ss},\,\,\rho^{th})$ provides a useful measure to
distinguish the steady state $\rho^{ss}$ from the thermal
state $\rho^{th}$ through quantum measurements \cite{trace distance2}.
Therefore, if $d(\rho^{ss},\,\,\rho^{th})$ increases with system
parameters we say that the system is farther away from
thermalization. Also, the thermal state $\rho^{th}$ is chosen
to be
$\rho^{th}=\exp[-\hbar\omega_{pol}(\sigma_{1}^{\dagger}\sigma_{1}+\sigma_{2}^{\dagger}\sigma_{2})/k_{B}T_{R}]$
up to a normalization factor tr$(\rho^{th})$.

\begin{figure}[h]
\begin{center}
%\vspace{10cm}
\subfigure[] {
\includegraphics[width=0.35\textwidth,height=0.3\textwidth]{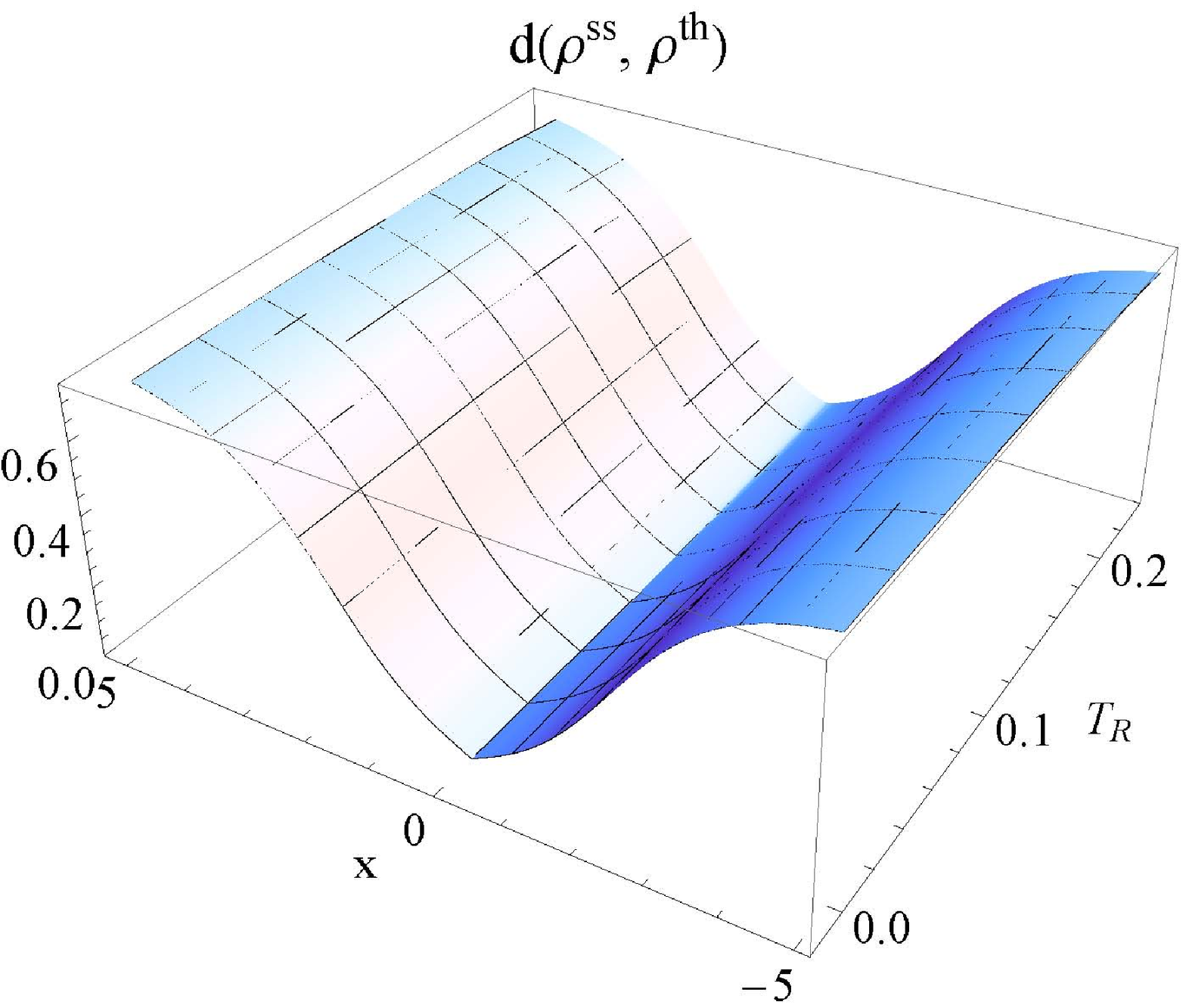}
\label{Fig.c2}} \hspace{20mm}\subfigure[] {
\includegraphics[width=0.35\textwidth,height=0.3\textwidth]{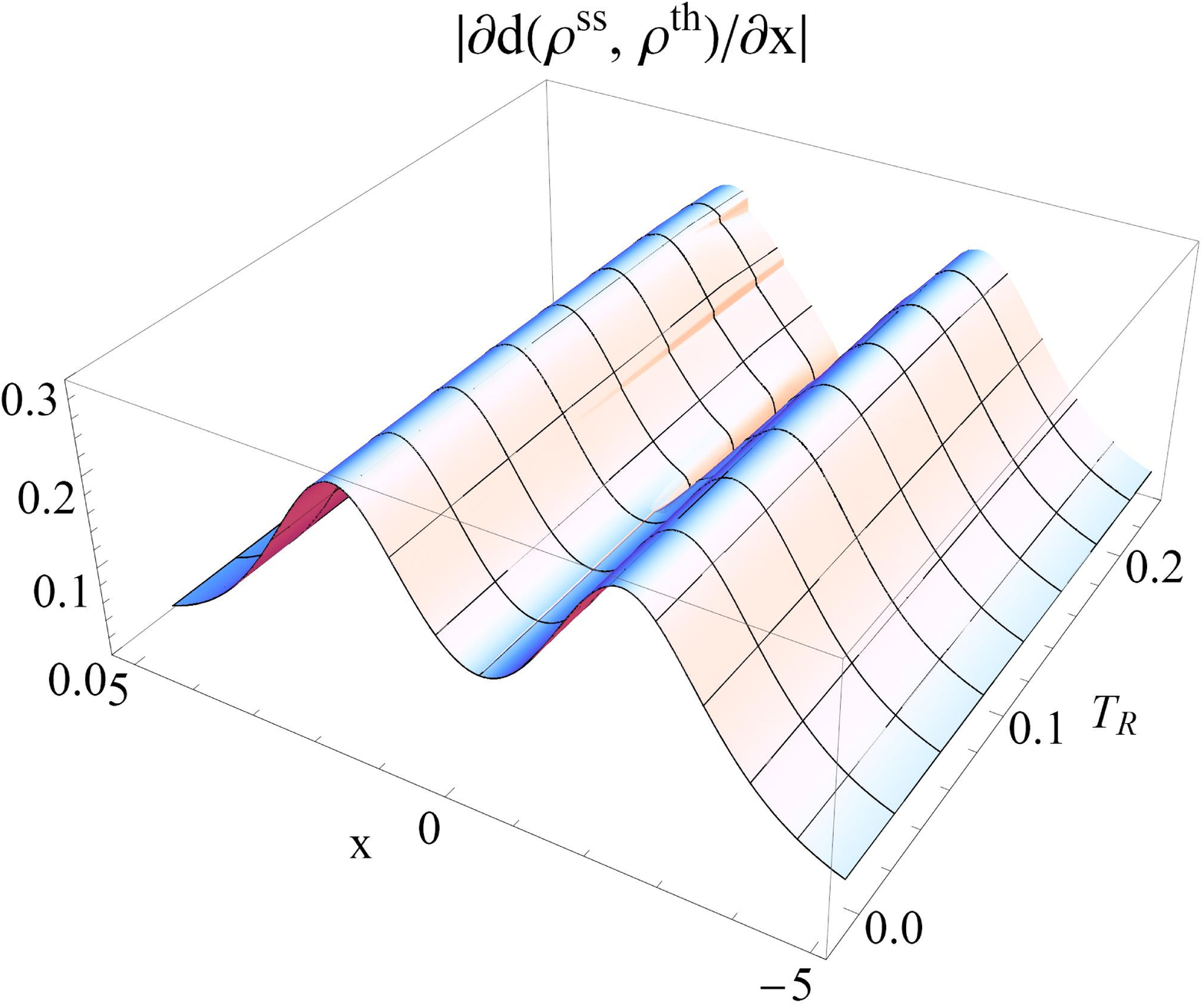}
\label{Fig.dc2} }
\end{center}
\caption{(color online).  The distance $d(\rho^{ss},\,\,\rho^{th})$
for (a) and the derivative $|\partial
d(\rho^{ss},\,\,\rho^{th})/\partial x|$ for (b) as functions of $x$
(proportional to the driving strength) and $T_{R}$.}
\end{figure}

Fig. \ref{Fig.c2} shows the distance $d(\rho^{ss},\,\,\rho^{th})$ as
a function of $x$ and $T_{R}$, where $x$ is a parameter defined in
Ref. \cite{two-state} (below Eq. (22)) and it is proportional to the
strength of the driving field. The relevant parameters $y=15,z=1.01$
(see Ref. \cite{two-state}). The unit of $T_{R}$ is
$\hbar\omega_{pol}/k_{B}$. It is seen in Fig. \ref{Fig.c2} that
the steady state is close to the thermal state if there is no
driving field, and for stronger driving field the steady state is
farther away from thermalization. This is reasonable from a physical
perspective as the driving field generally induces coherence (i.e.
non-zero off-diagonal elements in the polaritonic density matrix) for the
polaritons while the thermal state is diagonal. In addition, it
seems that $d(\rho^{ss},\,\,\rho^{th})$ does not depend on the
reservoir temperature. This may be because
$T_{R}\ll\hbar\omega_{pol}/k_{B}$ so that the effect of the thermal
agitation is rather small. The effect should certainly manifest
itself for larger $T_{R}$. However this regime is beyond the
approximation for the derivation of the effective master equation
($T_{R}\ll\hbar\omega_{pol}/k_{B}$) and it is in general not easily solvable
even with numerical calculations.

Comparing Fig. \ref{Fig.c2} for a fixed $T_{R}$ with the first plot
of Fig. 2 ($y=15$) in Ref. \cite{two-state}, one finds that they are
not consistent, especially for large $x$, for which
$d(\rho^{ss},\,\,\rho^{th})$ is very large while the polaritonic
entanglement is negligible. However, if one takes the derivative of
$d(\rho^{ss},\,\,\rho^{th})$ with respect to $x$, then a relationship appears. 
Fig. \ref{Fig.dc2} shows $|\partial
d(\rho^{ss},\,\,\rho^{th})/\partial x|$ as a function of $x$ and
$T_{R}$. It can be seen that there are two peaks for a fixed
temperature. This is similar to the first plot of Fig. 2 in Ref.
\cite{two-state}. Also the two plots are consistent for large $x$.
Therefore, it may be concluded that the change rate of the
thermalization with respect to the driving strength (rather than the
thermalization itself) is related to the polaritonic entanglement.
Physically, for a increase/decrease of the driving strength i.e.
more/less coherent energy is injected into the system, a more rapid
change of the thermal property (or the degree of thermalization) of
the system indicates that a stronger correlation (entanglement) is
established. The coherent energy refers to fact that  the driving field
induces off-diagonal elements in the polaritonic density matrix as
mentioned previously. One could conjecture that a more rapid change
of the degree of thermalization of the system may indicate that the
interaction between the two polaritons are stronger which leads to a
stronger entanglement between them.

\section{\large Conclusion}
In this paper, we show that long-distance steady state entanglement
in a lossy network of driven light-matter systems can be coherently
controlled through the tuning of the phase difference between the
driving fields. The role of driving phase field in engineering
interaction and entanglement in coupled atom-cavities was also
discussed in Ref. \cite{driving phase}. Here, it is found that in a
closed network of three-cavity-atom systems the maximum of
entanglement for any pair is achieved even when their corresponding
direct coupling is much smaller than their couplings to the third
party. This effect is reminiscent of coherent effects found in
quantum optics that coherent population transfers between otherwise
uncoupled levels through a third level using two classical coherent
fields. An alternative geometry: two-coupled cavities with three
driving fields is discussed. For finite temperature, we analyze the
thermalization of the two defect cavities coupled to one driven wave
guide. It is found that the change rate of the thermalization of the
system with respect to the driving strength (rather than the
thermalization itself) can indicate the degree of the polaritonic
correlation (entanglement).

Acknowledgement - This work was
supported by National Research Foundation \& Ministry of Education,
Singapore.  Li Dai would like to thank Dr. Jun-Hong An for
helpful discussions.

%%%%%%%%%%%%%%%%%%%%%%%%

\end{document}